\documentclass[prb,twocolumn,showpacs,preprintnumbers,amsmath,amssymb]{revtex4}

\usepackage{graphicx}
\usepackage{bbm}
\usepackage{float}

\begin{document}

\preprint{}
\title{Polar phonons and intrinsic dielectric response of the ferromagnetic insulating
spinel CdCr$_2$S$_4$  from first principles}
\author{Craig J. Fennie and Karin M. Rabe}
\affiliation{Department of Physics and Astronomy, Rutgers University,
        Piscataway, NJ 08854-8019}
\date{\today}

\begin{abstract}
We have studied the dielectric properties of the ferromagnetic spinel 
CdCr$_2$S$_4$ from first principles. Zone-center phonons and Born effective 
charges were calculated by frozen-phonon and Berry phase techniques within 
LSDA+U. We find that all infrared-active phonons are quite stable within the
cubic space group. The calculated static dielectric constant 
agrees well with previous measurements. These results suggest 
that the recently observed anomalous dielectric behavior in 
CdCr$_2$S$_4$ is not due to the softening of a polar mode. We suggest 
further experiments to clarify this point.

\end{abstract}


\maketitle


Multiferroics (MFs) displaying simultaneous ferroelectric (FE) and magnetic 
order are receiving considerable attention today.~\cite{spaldin.science.05} 
Basic questions on the nature of ferroelectricity required to coexist with
magnetism have been addressed,~\cite{hill.jpchemB.00} facilitating 
the search for new materials.~\cite{baettig.apl.05,azuma.jacs.05} Much effort
has been directed towards finding MFs displaying the magnetoelectric 
effect, due to both the desire to understand this particular fundamental 
manifestation of spin-lattice coupling and to the potential technological 
applications of controlling the magnetization (polarization) by an applied 
electric (magnetic) field.~\cite{hur.nature.04,fiebig.jpd.05}

One avenue for finding new MEs is to revisit known ferromagnetic (FM) 
insulators to look for ferroelectricity. CdCr$_2$S$_4$ is a member of a 
family of semiconducting FM chalcogenide $A$Cr$_2X_4$ spinels with $A$ = 
Cd, Hg and $X$ = S, Se.~\cite{baltzer.prl.65,nikiforov.crystal.99} The 
preponderance of evidence shows that CdCr$_2$S$_4$ crystallizes in the 
normal cubic spinel Fd3m space group over a wide temperature range, from
4K (lowest temperature measured) to decomposition temperature. CdCr$_2$S$_4$ 
is a simple Heisenberg ferromagnet with Cr$^{3+}$ spins ($S$=3/2) ordering 
at T$_c$=84K ($\Theta_{CW}$=155K).~\cite{baltzer.pr.151} The valence band 
consists of a relatively narrow Cr t$_{2g}$ peak hybridized with
mostly sulfur 2p states, while Cr e$_g$ states make up the lowest 
unoccupied states, across a gap of 1.6-1.8eV, in the conduction 
band.~\cite{miniscalco.prb.82,shanthi.jssc.00} Early interest in these materials 
was due to the observed coupling of the electronic structure and the lattice
to the magnetic 
subsystem.~\cite{harbeke.prl.66,gobel.jmmm.76,koshizuka.jap.78,wakamura.jap.88}
While the intrinsic nature of many of these effects has subsequently been 
questioned,~\cite{harbeke.ssc.70,capek.pss.77,koshizuka.prb.80,pouget.physicaB.92}
spin ordering has clearly been shown to have a relatively strong effect 
on the infrared-active and selected Raman-active phonon 
modes,~\cite{wakamura.jap.88,koshizuka.prb.80} making the chalcogenide spinels
an attractive system to study spin-lattice effects from first 
principles.~\cite{fennie.spin.phonon} Recently, CdCr$_2$S$_4$ was revisited 
to look for ferroelectricity, leading to the suggestion that CdCr$_2$S$_4$ is 
in fact a relaxor FE displaying a particularly large magnetocapacitive
effect.~\cite{hemberger.nature.05,lunkenheimer.05} A broad, frequency dependent
dielectric peak in the real part of the dielectric constant, of the type associated
with relaxor behavior, was observed. The dielectric constant at 10 Hz was shown
to change by 500$\%$ by application of a 5 T magnetic field, with the change 
decreasing rapidly with increasing excitation frequency. Finally, hysteresis 
loops indicating nonzero polarization (P$_s$) were observed below magnetic T$_c$ (well 
below the dielectric peak).  Octahedral-site Cr$^{3+}$ off-centering was proposed 
as a possible origin of this polar behavior on the basis of a previous 
analysis,~\cite{grimes.philmag.72} though such off-centering is unusual for magnetic 
ions.~\cite{hill.jpchemB.00} The relaxor behavior, associated in perovskites 
with polar nanoregions frustrated by chemical disorder,~\cite{samara.jpcm.03} 
was suggested to be due to geometrical frustration of local polar distortions 
which, as in the negative thermal expansion material ZrW$_2$O$_8$,~\cite{ramirez.mrs.05}
should result in a very-low-frequency polar phonon mode.

First-principles density functional methods have been highly successful in 
describing the structural, electronic, and magnetic properties of a 
variety of nonmagnetic and magnetic dielectrics, FEs,~\cite{waghmare.book.05} and 
multiferroics.~\cite{ederer.prb.05,neaton.prb.05,fennie.prb.05,seshadri.chemmat.01} 
First-principles methods have emerged as an ideal 
technique to differentiate between extrinsic and intrinsic properties of complex
dielectrics and FEs, e.g$.$ perovskite CaCu$_3$Ti$_4$O$_{12}$.~\cite{lixin.prb.02}
In this Letter we investigate the intrinsic FE and dielectric behavior of cubic 
spinel CdCr$_2$S$_4$ from first principles. We calculate all zone-center
phonons, Born effective charges and infrared-active (i.r$.$) mode oscillator 
strengths, allowing us to evaluate the static dielectric response of
a single-domain, stoichiometric, defect-free crystal and to evaluate the 
possibility of a soft-mode-driven ferroelectric instability or very-low-frequency 
geometrically-frustrated polar mode.

First-principles DFT calculations using PAW potentials were performed within 
LSDA and LSDA+U as implemented in {\sf VASP}.~\cite{VASP,PAW} The wavefunctions 
were expanded in plane waves up to a kinetic energy cutoff of 500 eV. Integrals 
over the Brillouin zone were approximated by sums on a $6 \times 6 \times 6$ 
$\Gamma$-centered $k$-point mesh; for density of states calculations this grid 
was increased to $14 \times 14 \times 14$. Phonon frequencies and 
eigendisplacements were calculated using the direct method where each symmetry
adapted mode~\cite{bilbao} was moved by approximately 0.01\AA. Born effective 
charge tensors were calculated by finite differences of the polarization using
the modern theory of polarization~\cite{king-smith.prb.93} as implemented in 
{\sf VASP}. All calculations were performed with the spin moments on the 
Cr-ions aligned ferromagnetically. 

Previous first-principles calculations~\cite{shanthi.jssc.00} have pointed 
out the inadequacy of LSDA for investigation of the electronic structure of 
CdCr$_2$S$_4$. In particular, LSDA fails to open a gap and a Cr 
t$_{2g}$ peak lies at the edge of the valence band. This failure, and its 
possible implications for the structural instabilities and dielectric 
response, can be addressed with LSDA+U.~\cite{anisimov.jpcm.97} While the 
value of U could be calculated from first principles with constrained LSDA 
calculations, we use the alternative phenomenologically
determined value U=3 eV, which approximately reproduces the dominant feature of 
photoemission data,~\cite{miniscalco.prb.82} namely a 
t$_{2g}$ peak 1.6 eV below the valence band.

\begin{table}[t]
\caption{ Structural, electronic, and magnetic properties of ferromagnetic CdCr$_2$S$_3$,
Space Group: Fd$\bar{3}$m.}
\begin{ruledtabular}
\begin{tabular}{lcccc}
& \multicolumn{2}{c}{Experiment}
& \multicolumn{2}{c}{Theory} \\ \hline

\begin{tabular}{l} structure\\$a$ ($\AA$)\\$u$\\electronic\\E$_g$ (eV) 
\\magnetic\\ $\mu$ ($\mu_b$)\\  \end{tabular}

&\begin{tabular}{c} 300K \\10.24\\0.390\\ \\1.57  \\ \\ \\  \end{tabular}

&\begin{tabular}{cc} 40K \\10.23\\0.390\\ \\1.8 \\ \\3\\ \end{tabular}

&\begin{tabular}{cc} \multicolumn{2}{c}{LSDA}
\\10.24&10.06\\0.392&0.391\\ \\0.77&0.00\\ \\2.84&2.80\\   \end{tabular}

&\begin{tabular}{c} U=3 eV
\\10.12\\0.390\\ \\1.47\\ \\2.88\\  \end{tabular}

\end{tabular}
\end{ruledtabular}
\label{table:lattice}
\end{table}


We performed a full relaxation of the lattice constant, $a$, and
anion internal parameter, $u$, both within LSDA and LSDA+U. We find that
the LSDA relaxed lattice constant, $a$=10.06~\AA, underestimates the 
experimental value by 1.8$\%$, slightly more than that typically ($<$1$\%$) 
found for LSDA calculations.  Within LSDA+U we find a value $a$=10.12\AA~ 
in much better agreement ($\sim$1$\%$) 
with the experimental value. This is consistent with other recent studies
for LaMnO$_3$~\cite{trimarchi.prb.05} 
and YMnO$_3$,~\cite{fennie.prb.05} and is worth noting, as the LSDA+U method 
has only recently begun to be applied to structural optimization. 

\begin{figure}[t]
\includegraphics[width=8.5cm,height=6.0cm]{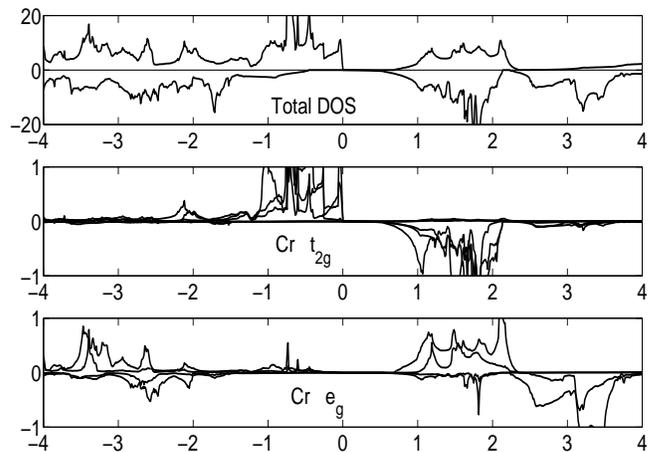}\\
\caption{\label{fig:FIG1} LSDA 
(Top) Total DOS and (Middle) site-projected DOS on the
Cr t$_{2g}$ and (Bottom) e$_g$ orbitals.}
\end{figure}

The electronic structure of the fully relaxed system within LSDA is similar to 
that of previous LMTO calculations.~\cite{shanthi.jssc.00} 
The total density-of-states (DOS) and site projected DOS for Cr 
t$_{2g}$ states is shown in Fig.~\ref{fig:FIG1} where a very small non-zero 
DOS, primarily t$_{2g}$ states of Cr, appears at the fermi level 
(note that this result is sensitive to volume where at the experimental 
lattice constant of $a$=10.24\AA~we find a gap even with LSDA of E$_g$=0.77 eV)
and the t$_{2g}$ orbitals of the valence band occupy states from
approximately 0.5 eV up to the Fermi level.
Taking U = 3.0 eV opens up a gap, E$_g$=1.5 eV, where the top valence band
is primarily oxygen p-states hybridized with Cr t$_{2g}$ (now pushed
down in energy $\sim$1.5eV) while the bottom
of the conduction band consists of Cr e$_g$ states with Cd s-states 
forming a wide conduction band.
We stress, given the ad hoc procedure used to obtain U, these results should only be
read as demonstrating that a reasonably good description of the electronic structure
of CdCr$_2$S$_4$ is possible within LSDA+U. Further studies should be performed
for quantitative features of the electronic structure. 
In both LSDA and LSDA+U, the calculated magnetic moments, $\mu$=2.80$\mu_b$ and 
$\mu$=2.9$\mu_b$ respectively, are slightly smaller than the 3$\mu_b$ expected for 
Cr$^{3+}$ at nominal valence . We summarized these results and compare with
experiment in Table~\ref{table:lattice}.

\begin{figure}[t]
\includegraphics[width=8.5cm,height=6.0cm]{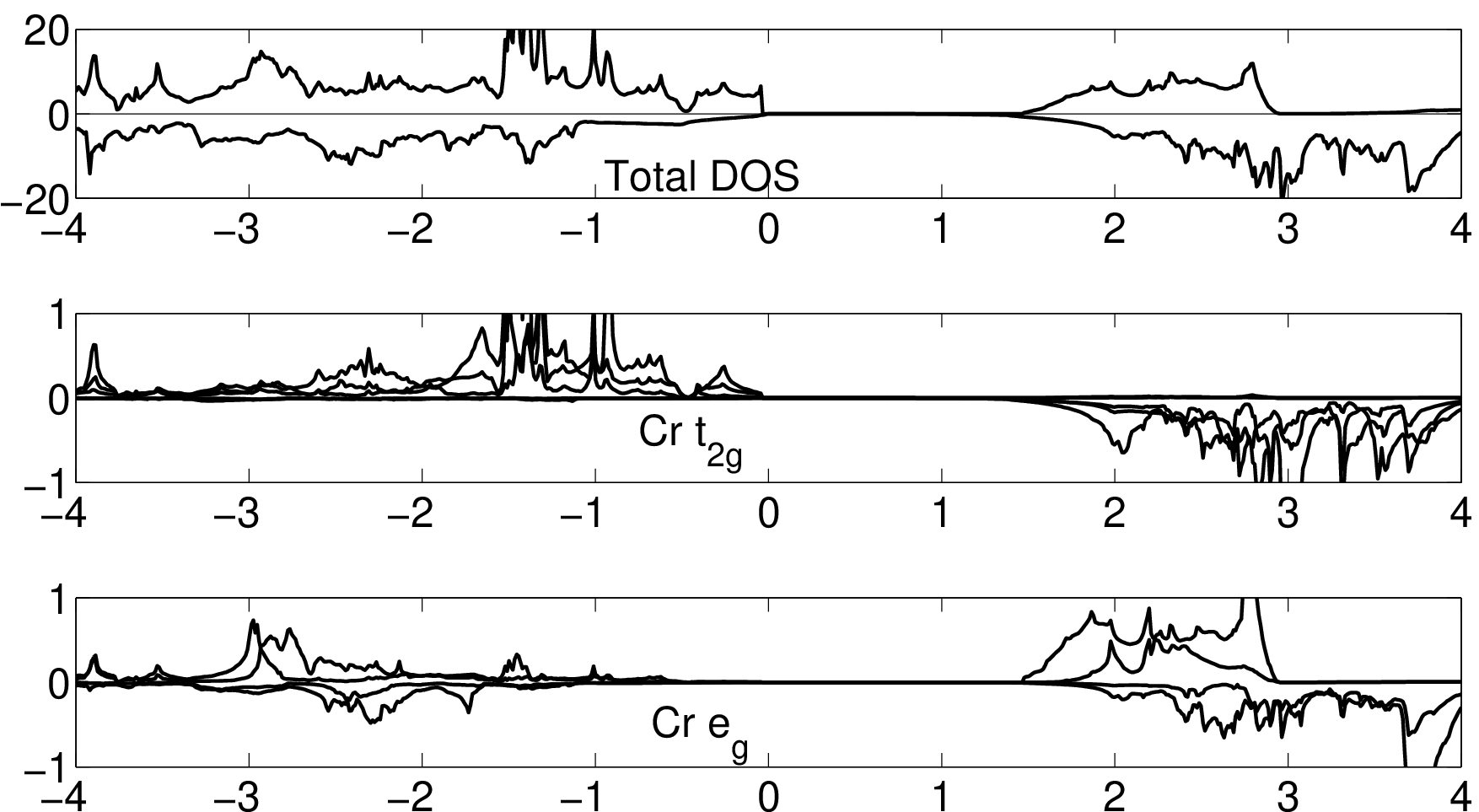}\\
\caption{\label{fig:FIG2} LSDA+U, U=3eV 
(Top) Total DOS and (Middle) site-projected DOS on the 
Cr t$_{2g}$ and (Bottom) e$_g$ orbitals.}
\end{figure}

\begin{table}[t]
\caption{ Infrared-active phonons frequencies, $\omega$ (cm$^{-1}$), effective
plasma frequencies, $\Omega_p$ (cm$^{-1}$), and dielectric constants of 
ferromagnetic CdCr$_2$S$_3$, Space Group: Fd$\bar{3}$m.}
\begin{ruledtabular}
\begin{tabular}{lcccc}
&Exp. Ref.~\onlinecite{lee.jap.71} & Exp. Ref.~\onlinecite{wakamura.jap.88} & \multicolumn{2}{c}{Theory} \\ \hline

\begin{tabular}{c} \\TO\\ 1\\2\\3\\4\\LO\\1\\2\\3\\4\\  $\epsilon_{phonon}$
\\$\epsilon_{\infty}$ \end{tabular}

&\begin{tabular}{ccc} 300K&\multicolumn{2}{c}{79K}\\$\omega$&$\omega$&$\Omega_p$\\ 
\\ \\322&327&400\\377&379&211\\ \\ \\ 
\\347&350\\390&393\\  1.9&1.8\\\multicolumn{2}{c}{7.8$\pm$0.2} \end{tabular}

&\begin{tabular}{cc} 300K&15K\\$\omega$&$\omega$ \\94&102\\239&249\\321&324\\378&379\\ 
\\98&105\\241&251\\349&352\\392&395\\  2.8&2.8\\8.0&7.6 \end{tabular}

&\begin{tabular}{cc} \multicolumn{2}{c}{U=3 eV}\\
$\omega$&$\Omega_p$\\104&74\\249&99\\339&400\\385&202\\ \\107\\251\\362\\398\\ 2.3\\ \\
 \end{tabular}

&\begin{tabular}{c} LSDA\\$\omega$
\\104\\247\\322\\366\\ \\106\\248\\349\\380\\ 2.5 \\ \\ \end{tabular}

\end{tabular}
\end{ruledtabular}
\label{table:ir}
\end{table}


Now that we have determined that a reasonably good description of the 
structural and electronic properties of CdCr$_2$S$_4$ can be 
obtained within LSDA+U, we proceed to our main topic, the lattice dynamics
and dielectric behavior. In Table~\ref{table:ir} we show our calculated
LSDA+U i.r$.$ active TO and LO modes and compare with those extracted from 
reflectivity experiments above and below the FM transition. We find that 
all i.r$.$ modes are stable and in fact are quite hard, with the lowest 
frequency at 104 cm$^{-1}$. Agreement of our calculated phonon 
frequencies with experiment is within 2$\%$, except for mode TO3 which 
still only differs by $\approx$ 3.5$\%$. To further test the quality of our 
calculations we computed the Raman active phonons, Table~\ref{table:raman}, 
where again agreement with experiment is within a percent. This is 
even better than typical for first-principles phonon calculations, where
agreement to 1 or 2 wavenumbers should be considered fortuitous. 
To check the possible effect of the underestimation of the volume on 
the phonon frequencies, we repeated the calculations at the experimental volume 
finding very minor changes (e.g. the lowest frequency i.r$.$ mode is now 
at 102 cm$^{-1}$). 
While non-zero U had a noticeable effect on the structural parameters, we
find that the effect of non-zero U on the phonon frequencies 
is minimal (with an overall improved agreement with the experimental values)
compared with the LSDA, also shown in Table~\ref{table:ir}.
Therefore, LSDA+U provides a consistent theory to compute the lattice, 
electronic, and magnetic structures of CdCr$_2$S$_4$.

The intrinsic static dielectric response is calculated as 
$$\epsilon_{0} = \epsilon_{\infty} + \sum_{m} {\Omega_{p,m}^2 \over \omega_{m}^2} $$
where $\epsilon_{\infty}$ is the electronic dielectric constant and 
the second term on the r.h.s$.$, $\epsilon_{ph}$, is the sum of 
contributions from each i.r$.$-active phonon, called the oscillator 
strength. $\Omega_{p,m}$ and $\omega_{m}$ are the effective plasma 
frequency and i.r$.$ phonon frequency for mode $m$, respectively. 
A large dielectric response can be the result of the presence of one 
or more very low frequency polar phonons 
and/or
anomalously large effective plasma frequencies. Although $\epsilon_{\infty}$ 
can be calculated from first principles, most readily with density 
functional perturbation theory,~\cite{gonze.prb.97,baroni.rmp.01} 
our main interest here is the phonon contribution and thus we use
the experimental value $\epsilon_{\infty}$ $\approx$ 8.
We computed $\Omega_{p}$ within LSDA+U from calculations of the Born 
effective charge tensors, which were found to be close to the nominal charges
and the real-space eigendisplacements of the phonons.~\cite{gonze.prb.97}
The results are shown in Table~\ref{table:ir} and compared with available 
experimental values, where again we find excellent agreement. We see that the 
most polar mode (TO3), i.e$.$ the mode with the largest $\Omega_{p}$, is at 
$\approx$ 300 cm$^{-1}$, thus reducing its oscillator strength $\mathcal{O}$(1). 
The lower frequency modes, TO1 and TO2, are 
weakly polar, as reflected by the very small value of $\Omega_{p}$, which may be why 
Ref.~\onlinecite{lee.jap.71} classified these modes as non-polar.
Also note, the relative value of $\Omega_{p}$, low for TO1 and TO2, high for TO3
and TO4, is due to the fact that the eigendisplacements of these modes are 
qualitatively different. TO1 and TO2 modes consist mainly of Cd$^{2+}$ ions moving 
against both Cr$^{3+}$ and S$^{2-}$ ions while TO3 and TO4 are dominated by
Cr$^{3+}$ moving against S$^{2-}$. We point out that this qualitative 
difference of the eigendisplacements is also responsible for the mode-dependent
phonon anomalies observed at T$_c$, which we will explore in a future 
publication.~\cite{fennie.spin.phonon}
The total contribution to the phonon part of the calculated dielectric constant 
$\epsilon_{ph}$=2.3, agrees well with that extracted from 
reflectivity measurements, $\epsilon_{ph}$=2.8.~\cite{wakamura.jap.88}  
Note that $\epsilon_{ph}$=1.9 measured in Ref.~\onlinecite{lee.jap.71} 
only contains contributions of TO3 and TO4. This should be compared with our
calculations $\epsilon$(TO3)+$\epsilon$(TO4)=1.7.  
Combining $\epsilon_{ph}$ with the experimental value of $\epsilon_{\infty}$,
the computed intrinsic value of the static dielectric constant of CdCr$_2$S$_4$, 
$\epsilon_0$, is found to be $\approx$10. This is completely consistent with
the values obtained by reflectivity measurements and
the capacitor measurements at 3 GHz of 
Ref.~\onlinecite{hemberger.nature.05,lunkenheimer.05}.

\begin{table}[t]
\caption{ Raman-active phonons frequencies, cm$^{-1}$, of ferromagnetic CdCr$_2$S$_3$,
Space Group: Fd$\bar{3}$m.}
\begin{ruledtabular}
\begin{tabular}{lccc}
& Exp. Ref.~\onlinecite{koshizuka.prb.80} & \multicolumn{2}{c}{Theory} \\ \hline

\begin{tabular}{l} \\T$_{2g}$(1)\\E$_{g}$\\T$_{2g}$(2)\\
T$_{2g}$(3)\\A$_{1g}$   \end{tabular}

&\begin{tabular}{cc} 300K&40K \\101&105\\256&257\\280&281\\351&353\\394&396 \end{tabular}

&\begin{tabular}{c} LSDA
\\99\\257\\280\\352\\394   \end{tabular}

&\begin{tabular}{c} U= 3 eV
\\100\\262\\282\\352\\393  \end{tabular}

\end{tabular}
\end{ruledtabular}
\label{table:raman}
\end{table}


Our first-principles calculations of the phonons and dielectric 
response clearly show that CdCr$_2$S$_4$ has neither a FE instability 
nor very-low-frequency polar modes which might be indicative of geometric
frustration.~\cite{hancock.prl.04} In fact, the intrinsic dielectric 
response of CdCr$_2$S$_4$ is found to be that of a normal dielectric.
Further, the lack of a zone-center FE instability implies that no 
P$_s$ is expected at low temperatures. These results
are in direct contrast to previous first-principles calculations of perovskite 
relaxors (on ordered supercells) where FE instabilities and anomalously large Born 
effective charges, yielding large oscillator strengths for certain modes, are 
found.~\cite{choudhury.prb.05,prosandeev.prb.04}

We considered the possibility of other structural instabilities, 
in particular a Cr-off-centering antiferroelectric transition 
to the $F\bar{4}3m$ space group. In our 14-atom unit cell, this would 
be produced by the freezing in of a silent zone-center A$_{2u}$ phonon.
The possibility of this phase transition of this type has been the subject
of previous speculation (but never experimentally
demonstrated) for many spinels.~\cite{schmid.jpc.74,grimes.philmag.72}  
Our calculation reveals that both A$_{2u}$ modes are quite hard with 
frequencies greater than 300 cm$^{-1}$. In addition, the frequencies 
of the other silent zone-center phonon modes, E$_u$, T$_{1g}$, and T$_{2u}$, 
were all calculated to be greater than 100 cm$^{-1}$.

Then what could be the origin of the anomalous dielectric response reported
for CdCr$_2$S$_4$? While our results clearly rule out soft polar modes, one
possibility might be the coupling of a polar mode to a zone-boundary mode, as 
we previously have suggested happens for YMnO$_3$ (note, unlike that of YMnO$_3$, 
symmetry would require this transition to be first order). A related mechanism
might involve coupling to an incommensurate structural distortion.
Although most previous experiments are inconsistent with such a picture
we suggest that detailed structural experiments should be performed. From the
theoretical side, first-principles calculations of the phonons throughout 
the Brillouin zone will provide further insight to this type of scenario.
Another possibility our work suggests is that the dielectric response 
and polarization observed in Ref.~\onlinecite{hemberger.nature.05,lunkenheimer.05}
may not be intrinsic.~\cite{lixin.prb.02} While experimentally the issue of contacts
have been explored, less has been done (to our knowledge) to characterize
the defect structure.~\cite{cohen.jap.03} This is important given the recent 
claims that the relaxor behavior disappears in annealed single crystals and 
polycrystalline samples.~\cite{hemberger.july.05} We suggest controlled defect 
experiments be performed to gain a more precise understanding of this effect.

To summarize, the computed dielectric constant of 10 is consistent with
the measurements at frequencies $\mathcal{O}$(GHz), while no i.r$.$ active
phonon of the cubic spinel structure can account for the larger responses
observed at lower frequencies. This suggests that the origin for the observed
relaxor behavior in CdCr$_2$S$_4$ is something other than a displacive polar soft mode.

Useful discussions with S-W$.$ Cheong  and  D.H$.$ Vanderbilt 
are acknowledged. This work was supported by NSF-NIRT Grant No. DMR-0103354.
CF acknowledges support of Lucent/Bell Labs-Rutgers Fellowship.


\end{document}